%% file: pw14_wohlfeil_arxiv.tex
\documentclass[a4paper]{spie}  %>>> use this instead for A4 paper
\usepackage[]{graphicx}
\usepackage{amssymb,amsmath,psfrag,url}
\usepackage{verbatim}
\usepackage{epstopdf}
\usepackage[config]{subfig}

\title{
Numerical Simulation of Grating Couplers\\ for Mode Multiplexed Systems
}

\author{
Benjamin Wohlfeil,\supit{\,a}
Sven Burger,\supit{\,bc}
Christos Stamatiadis,\supit{\,a} 
Jan Pomplun,\supit{\,c} \\
Frank Schmidt,\supit{\,bc} 
Lars Zimmermann,\supit{\,d} 
Klaus Petermann\supit{\,a} 
\skiplinehalf
\supit{a}
Fachgebiet Hochfrequenztechnik, TU Berlin,
Einsteinufer~25,
D\,--\,10\,587 Berlin,
Germany
\smallskip\\
\supit{b}
Zuse Institute Berlin\,(ZIB),
Takustrasse~7,
D\,--\,14\,195 Berlin,
Germany
\smallskip\\
\supit{c}
JCMwave GmbH,
Bolivarallee~22, 
D\,--\,14\,050 Berlin,
Germany
\smallskip\\
\supit{d}
IHP GmbH,
Im Technologiepark~25
D\,--\,15\,232 Frankfurt (Oder),
Germany
}
\authorinfo{
Corresponding author: B. Wohlfeil\\
URL: http://www.tu-berlin.de\\
URL: http://www.zib.de\\
URL: http://www.jcmwave.com\\
URL: http://www.ihp-microelectronics.com
}

\begin{document}
\maketitle
%\today

%%%%%%%%%%%%%%%%%%%%%%%%%%%%%%%%%%%%%%%%%%%%%%%%%%%%%%%%%%%%%%%%%%%%%%%%%%%%%%%%%%%%%%%%%%%%%%%%%%%%
%%%%%%%%%%%%%%%%%%%%%%%%%%%%%%%%%%%%%%%%%%%%%%%%%%%%%%%%%%%%% 
%% SPIE Copyright form 
\noindent
This paper will be published in Proc.~SPIE Vol.~{\bf 8988}
(2014) 89880K, ({\it Integrated Optics: Devices, Materials, and Technologies XVIII}, DOI: 10.1117/12.2044461), 
and is made available 
as an electronic preprint with permission of SPIE,  \includegraphics[scale=0.015]{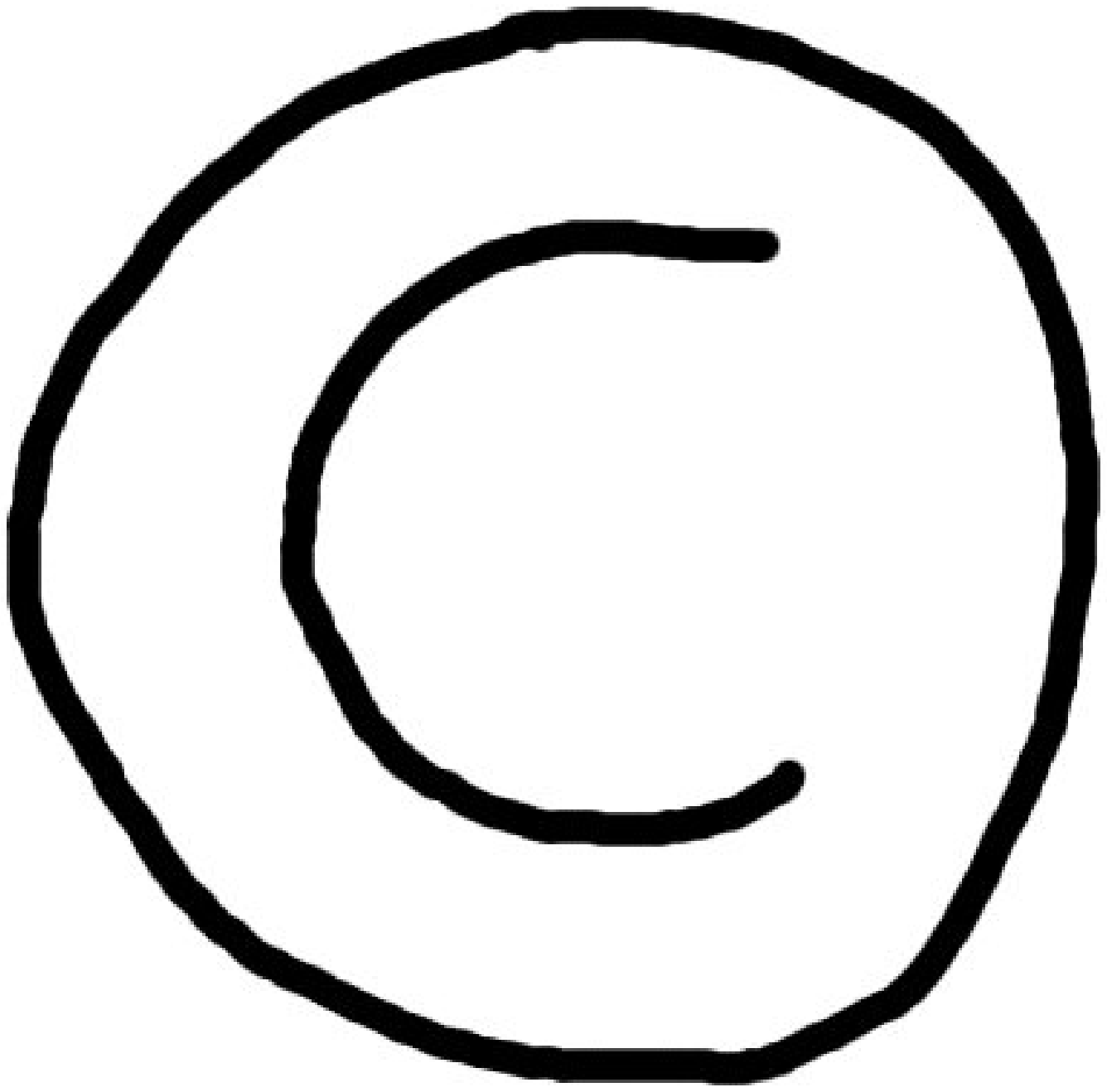} Society of Photo-Optical
Instrumentation Engineers (2014).
One print or electronic copy may be made for personal use only. 
Systematic reproduction and distribution, duplication of any 
material in this paper for a fee or for commercial purposes, 
or modification of the content of the paper are prohibited.
%Please see original paper for images at higher resolution. 

%%%%%%%%%%%%%%%%%%%%%%%%%%%%%%%%%%%%%%%%%%%%%%%%%%%%%%%%%%%%%%%%%%%%%%%%%%%%%%%%%%%%%%%%%%%%%%%%%%%%
\begin{abstract}
A numerical investigation of a two dimensional integrated fiber grating coupler capable of exciting several LP fiber modes in both TE and TM polarization is presented. Simulation results and an assessment of the numerical complexity of the 3D, fully vectorial finite element model of the device are shown.
\end{abstract}


%%%%%%%%%%%%%%%%%%%%%%%%%%%%%%%%%%%%%%%%%%%%%%%%%%%%%%%%%%%%%%%%%%%%%%%%%%%%%%%%%%%%%%%%%%%%%%%%%%%%
%%%%%%%%%%%%%%%%%%%%%%%%%%%%%%%%%%%%%%%%%%%%%%%%%%%%%%%%%%%%%%%%%%%%%%%%%%%%%%%%%%%%%%%%%%%%%%%%%%%%
\section{Introduction}
To avoid the approaching capacity crunch in single mode fiber based transmission systems, which are slowly coming to their theoretical limits, much research has been focused recently on the exploitation of spatial division multiplexing (SDM), especially in the form of multimode transmission. However, to compete with current technologies, the additional modes have to be generated and received in an efficient and cost effective way. Current solutions either lack efficiency \cite{salsi11} or are too complex for a cost effective implementation \cite{klaus12}. Integration on the other hand  has proven itself in the past to be a big improvement in these areas. Therefore several solutions based on photonic  integrated circuits have been presented \cite{doerr11,fontaine12}, albeit still with room for optimization. The device presented in this paper aims to improve upon other solutions by using a single standard sized two dimensional fiber grating coupler, which couples the commonly used fiber modes LP$_{01}$, LP$_{11a}$, LP$_{11b}$ and LP$_{21a}$ in TE and TM polarization from photonic nanowires on silicon on insulator to a standard few mode fiber with a high modal overlap and low losses. Simulation results for the performance of the device as well as an estimation of the numerical effort for optimization are presented.

The paper is structured as follows: First the concept for the device itself is presented, followed by an overview over the numerical effort required for the modelling of the device in a three dimensional, fully vectorial finite element solver. Simulation results for the theoretical performance of the coupler are shown and finally the results are summarized with an outlook to future work.

\begin{figure}
  \centering
  \includegraphics[scale=0.75]{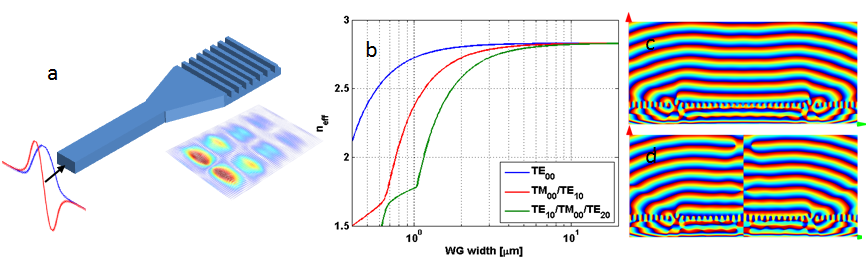}
  \caption{a) Fiber grating coupler excited with either TE$_{00}$ (blue) or TE$_{10}$ mode (red) together with the scattered field of the TE$_{10}$ mode. b) Effective refractive indices of the first three guides modes in nanowires on SOI as a function of waveguide width at $\lambda$ = 1550\,nm. c) and d) Phase profiles of the electric fields in a cross section of a fiber grating coupler excited from both ends with (d) and without (c) phase shift (pseudo-color plots: blue to red corresponds to phase $-\pi$ to $\pi$).}
  \label{fig:fig1}
\end{figure}

%%%%%%%%%%%%%%%%%%%%%%%%%%%%%%%%%%%%%%%%%%%%%%%%%%%%%%%%%%%%%%%%%%%%%%%%%%%%%%%%%%%%%%%%%%%%%%%%%%%%
%%%%%%%%%%%%%%%%%%%%%%%%%%%%%%%%%%%%%%%%%%%%%%%%%%%%%%%%%%%%%%%%%%%%%%%%%%%%%%%%%%%%%%%%%%%%%%%%%%%%
\section{Concept}
For the excitation of fiber modes the intensity and phase profile of the scattered field of the grating coupler have to be matched to the fiber mode. A standard fiber grating coupler only couples to the LP$_{01}$ mode, since it only requires a single intensity maximum with a flat phase. For higher order modes, however, several intensity maxima with a relative phase shift between them have to be generated. In the case of the first higher order fiber mode (LP$_{11a}$), the TE$_{10}$ integrated waveguide mode can be used to provide the necessary field distribution since it already comes with two intensity maxima with a relative phase shift of 180$^\circ$ between them. According to the Bragg condition a grating coupler capable of scattering such a mode can be designed if the period $\Lambda$ is chosen in accordance with the phase constant $\beta_{TE10}$ of the TE$_{10}$ mode in the integrated waveguide.

In Figure~\ref{fig:fig1}a the scattered vector field above such a grating is shown. As can be seen, the grating excites two very distinct intensity maxima along the transversal direction of the waveguide as it is prescribed by the TE$_{10}$ waveguide mode. In the lateral direction the grating shows an identical behavior to a standard fiber grating coupler for the fundamental waveguide mode.

For large waveguide widths of around 10\,$\mu$m and larger, which is the usual dimension for fiber grating couplers, the effective indices of the first few guided modes in nanowires are approximately equal (c.f.~Figure ~\ref{fig:fig1}b), resulting in - according to the Bragg relation - equal scattering characteristics, when diffracted by a grating with fixed grating period $\Lambda$. Indeed, simple scattering simulations show equal diffraction angle and fraction of out-coupled power for one grating fed with TE$_{00}$ and TE$_{10}$ mode, respectively. Therefore, analogously to the excitation of the fundamental fiber mode LP$_{01}$ through the fundamental waveguide mode TE$_{00}$ the first higher order waveguide mode TE$_{10}$ can be used to excite a LP$_{11}$ fiber mode with a single grating coupler.
For the generation of the other LP$_{11}$ fiber mode the grating is fed from both ends with the fundamental waveguide mode TE$_{00}$ with a relative phase difference of 180$^\circ$ between them. By doing this two distinct intensity maxima along the lateral waveguide direction are excited, as it is necessary for the LP$_{11b}$ fiber mode. If on the other hand the phase shift between the incoming TE$_{00}$ modes is 0$^\circ$ the fields scattered from the opposing ends of the grating do not feature the necessary phase difference to maintain two distinct intensity maxima, resulting in a degeneration to a single intensity maximum. This effect can again be utilized to excite the fundamental fiber mode LP$_{01}$. Figures \ref{fig:fig1} c and d show the phase profiles of the electric fields in a cross-sectional view of a grating coupler that is excited from both ends (left and right) with the fundamental waveguide mode with a 0$^\circ$ (c) and a 180$^\circ$ (d) phase difference, respectively.

\begin{figure}[hbt]
  \centering
  \includegraphics[scale=0.75]{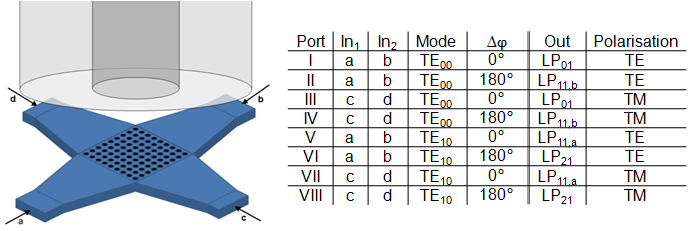}
  \caption{Schematic of the proposed device together with a table of all excitable modes.}
  \label{fig:fig2}
\end{figure}

Since the grating shows equal scattering properties for  TE$_{00}$ and TE$_{10}$ modes the same is true for an incoming TE$_{10}$ from both ends of the grating: If the phase difference is 0$^\circ$ the fields from both ends will melt together, producing the LP$_{11a}$ fiber mode. With a phase difference of 180$^\circ$ the incoming TE$_{10}$ modes from both ends are detached from each other creating four distinct intensity maxima as exhibited by the LP$_{21}$ fiber mode. The complete layout of the grating together with a table of all excitable fiber modes can be seen in Figure~\ref{fig:fig2}. Note that a two dimensional grating is used for the operation in both TE and TM polarization. To excite fiber modes in TE polarization input waveguides a and b are used, while for TM polarized modes waveguides c and d need to be excited. Each waveguide pair is fed with TE$_{00}$ or TE$_{10}$ modes with either 0$^\circ$ or 180$^\circ$ phase difference resulting in four different fiber modes per polarization. Due to the forced symmetry by the way of excitation the coupled fiber needs to be strictly vertical in contrast to conventional fiber-grating coupling. Therefore, more care has to be taken, when considering reflections between fiber facet and grating.

Since the orthogonality of the fiber modes is maintained at every part of the device all modes can be excited independently of each other at the same time.

%%To obtain a sufficiently large distance between the intensity maxima to ensure a large field overlap between the scattered field %%and the LP$_{11b}$ fiber mode some design optimization has to be carried out. The two most important parameters here are the %%number of periods and the etch depth of the grating coupler, which together determine the relative distance distance between %%intensity maxima.

\section{Numerical method}
\input{pw14_numerical_method.tex}

\section{Numerical results}
 In the best case one can achieve a consistently high field overlap between scattered fields and desired fiber modes together with a high directionality (fraction of power scattered towards the fiber) of the grating. For this device an epitaxial silicon overlay of 150\,nm on top of the grating was employed to increase the overall coupling efficiency of the grating. The two most important design parameters for this device are the grating length as determined by the number of grating periods and the etch depth of the grating, which mainly defines the grating strength and therefore at which position along the axis of propagation in the grating the intensity maxima are located. Together the parameters need to be chosen so that the distance between fields scattered from opposing ends of the grating is not too big to prevent a melting together of the fields for the generation of the LP$_{01}$ mode and not too small for the formation of the LP$_{11b}$ fiber mode. The shape of the LP$_{11a}$ mode is mainly determined by the width of the grating coupler, which is equal to its length, due to symmetry (see Figure~\ref{fig:fig2}). The grating period is for the most part a function of the wavelength the device is operated at. The duty cycle of the grating (fraction of etched to non-etched area) can be chosen freely, but should not be too small since this would result in a transparent grating.
Here a grating etch depth of 210\,nm was chosen, since in combination with a silicon overlay of 150\,nm on top of the 220\,nm waveguide height results in a high directionality (fraction of power scattered upwards) of the grating coupler. Together with 21 grating periods this leads to good field overlaps between scattered fields and fiber modes. As a figure of merit across all fiber modes $\rho = p \cdot \eta_{LP01} \cdot \eta_{LP11a} \cdot \eta_{LP11b}$ was introduced, where p is the fraction of power scattered towards the fiber independent of the way of excitation and $\eta_i$ is the overlap with the excited fiber mode. The overlap with LP$_{21}$ is missing here since only one of the modes in this group can be excited and therefore the mode group would be incomplete.

\begin{figure}[ht]
  \centering
  \includegraphics[scale=0.75]{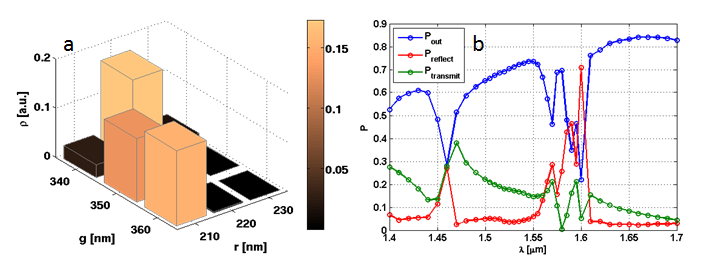}
  \caption{a) Measure of grating efficiency as a function of two design parameters g and r at $\lambda$=1550\,nm. b) Spectral performance of a 2D grating with out-coupled power (blue), reflected power (red) and transmitted power (green).}
  \label{fig:fig3}
\end{figure}

Figure \ref{fig:fig3}a shows the figure of merit $\rho$ as a function of two grating design parameters groove-width g (etched part of the grating period) and ridge-width r (non etched part of the grating period). A grating with a period of around 560\,nm gives the best results on average across all three fiber modes. However, for slightly varying grating periods the performance of the device drops rapidly, indicating a low tolerance for the manufacturing process. Nevertheless, with an adaptation of the etch depth a potential deviation of the grating period can be compensated. A filter curve of a two dimensional grating with a period of 560\,nm is depicted in Figure~\ref{fig:fig3}b, where the blue line indicates the out-coupled power. Due to the silicon overlay and the deep etch depth such a grating can be quite efficient, peaking at a wavelength of 1550\,nm. For longer wavelengths the power scattered towards the fiber decreases sharply and transmission (green line) and reflection (red line) increase.

\begin{figure}[ht]
  \centering
  \includegraphics[scale=0.75]{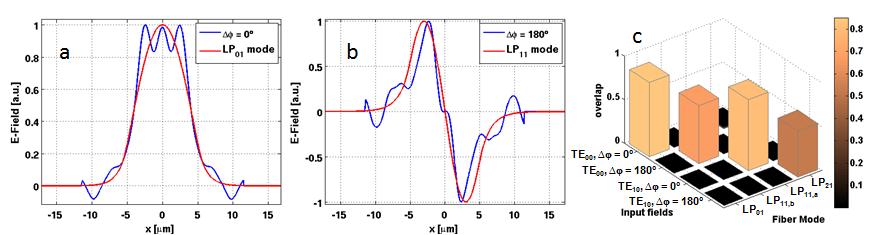}
  \caption{a) Cross section of the electric field of a 2D grating excited from both ends without phase difference (blue) and fiber mode LP$_{01}$. b) Electric field with phase difference of 180$^\circ$ (blue) and fiber mode LP$_{11b}$ (red). c) Overlap integrals for all combinations of excitation of the grating and fiber modes.}
  \label{fig:fig4}
\end{figure}

A cross section of the electric fields emitted by the grating can be seen in Figures~\ref{fig:fig4}a and b. 
The overlap with the desired fiber modes (here LP$_{01}$ and LP$_{11b}$) is clearly evident, 
thus -- together with the high directionality of the grating -- giving a good coupling efficiency. 
Figure \ref{fig:fig4}c shows the complete overlaps calculated from the fields of fully 3D light scattering simulations 
with all excitable fiber modes. 
Being in a not fully excitable mode group the performance of the LP$_{21a}$ mode was neglected for the 
optimization of the grating, resulting in the lowest field overlap of only 50\%. 
All other modes show a high overlap of 70\% and above with the scattered fields, when the grating 
is excited correspondingly. 
Higher overlaps of 90\% for singular modes can be achieved with adapted grating designs, but the 
performance of other modes would be affected negatively. 
Due to the mentioned orthogonality of the fields involved at every stage in the device the extinction 
of unwanted modes is also very large, with power ratios always below 10$^{-6}$.

%%%%%%%%%%%%%%%%%%%%%%%%%%%%%%%%%%%%%%%%%%%%%%%%%%%%%%%%%%%%%%%%%%%%%%%%%%%%%%%%%%%%%%%%%%%%%%%%%%%%
%%%%%%%%%%%%%%%%%%%%%%%%%%%%%%%%%%%%%%%%%%%%%%%%%%%%%%%%%%%%%%%%%%%%%%%%%%%%%%%%%%%%%%%%%%%%%%%%%%%%
%%%%%%%%%%%%%%%%%%%%%%%%%%%%%%%%%%%%%%%%%%%%%%%%%%%%%%%%%%%%%%%%%%%%%%%%%%%%%%%%%%%%%%%%%%%%%%%%%%%%
\section{Conclusion}
We presented a numerical investigation of a two dimensional fiber grating coupler capable of 
exciting the LP$_{01}$, LP$_{11a}$, LP$_{11b}$ and LP$_{21a}$ fiber mode in TE and TM polarization. 
Accurate solutions for the  challenging 3D light scattering problem have been validated in a 
convergence study. 
Simulations show a good performance of the grating and high modal overlaps between scattered fields 
and fiber modes and a very low crosstalk between modes.

%%%%%%%%%%%%%%%%%%%%%%%%%%%%%%%%%%%%%%%%%%%%%%%%%%%%%%%%%%%%%%%%%%%%%%%%%%%%%%%%%%%%%%%%%%%%%%%%%%%%
%%%%%%%%%%%%%%%%%%%%%%%%%%%%%%%%%%%%%%%%%%%%%%%%%%%%%%%%%%%%%%%%%%%%%%%%%%%%%%%%%%%%%%%%%%%%%%%%%%%%
\section*{Acknowledgments}
This work has been partially supported 
by the European commission through project ICT-GALACTICO, grant agreement 258407,
by BMBF within project {\sc Mosaic} (FKZ 13N12438),  
and by German Research Foundation DFG through SFB 787.

\bibliography{pw14_bibliography}
\bibliographystyle{spiebib}

%%%%%%%%%%%%%%%%%%%%%%%%%%%%%%%%%%%%%%%%%%%%%%%%%%%%%%%%%%%%%%%%%%%%%%%%%%%%%%%%%%%%%%%%%%%%%%%%%%%%

\end{document}

%% file: pw14_numerical_method.tex
%\subsection{Finite-element method}
For optimizing the design parameters of a 
fiber grating coupler we perform numerical simulations of light propagation through 
the device. 
The relevant physical model in this case are Maxwell's equations, with the  
full 3D geometrical structure of the device. 
As we are interested in the wavelength-dependence of the coupling efficiency of 
specific field patterns we perform time-harmonic (steady-state) light scattering simulations 
of incoming waveguide modes for the wavelength range of interest. 
For this we use the Maxwell solver JCMsuite~\cite{Burger2008ipnra} based on the 
finite-element method (FEM)~\cite{Pomplun2007pssb}. 

\begin{figure}[b]
\centering
\subfloat[]{\label{fig_conv_t}
\includegraphics[width=0.49\textwidth]{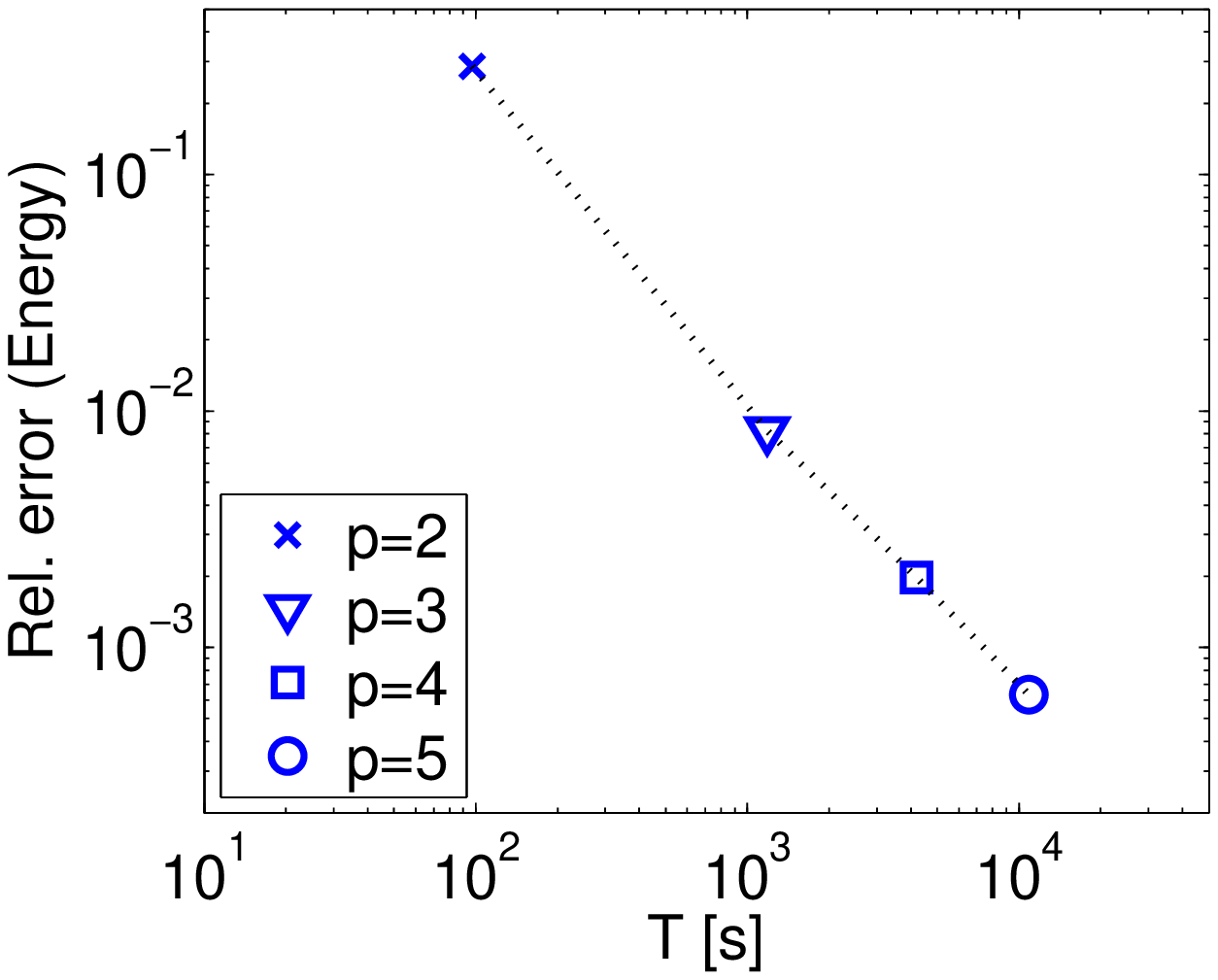}}
\hfill
\subfloat[]{\label{fig_conv}
\includegraphics[width=0.49\textwidth]{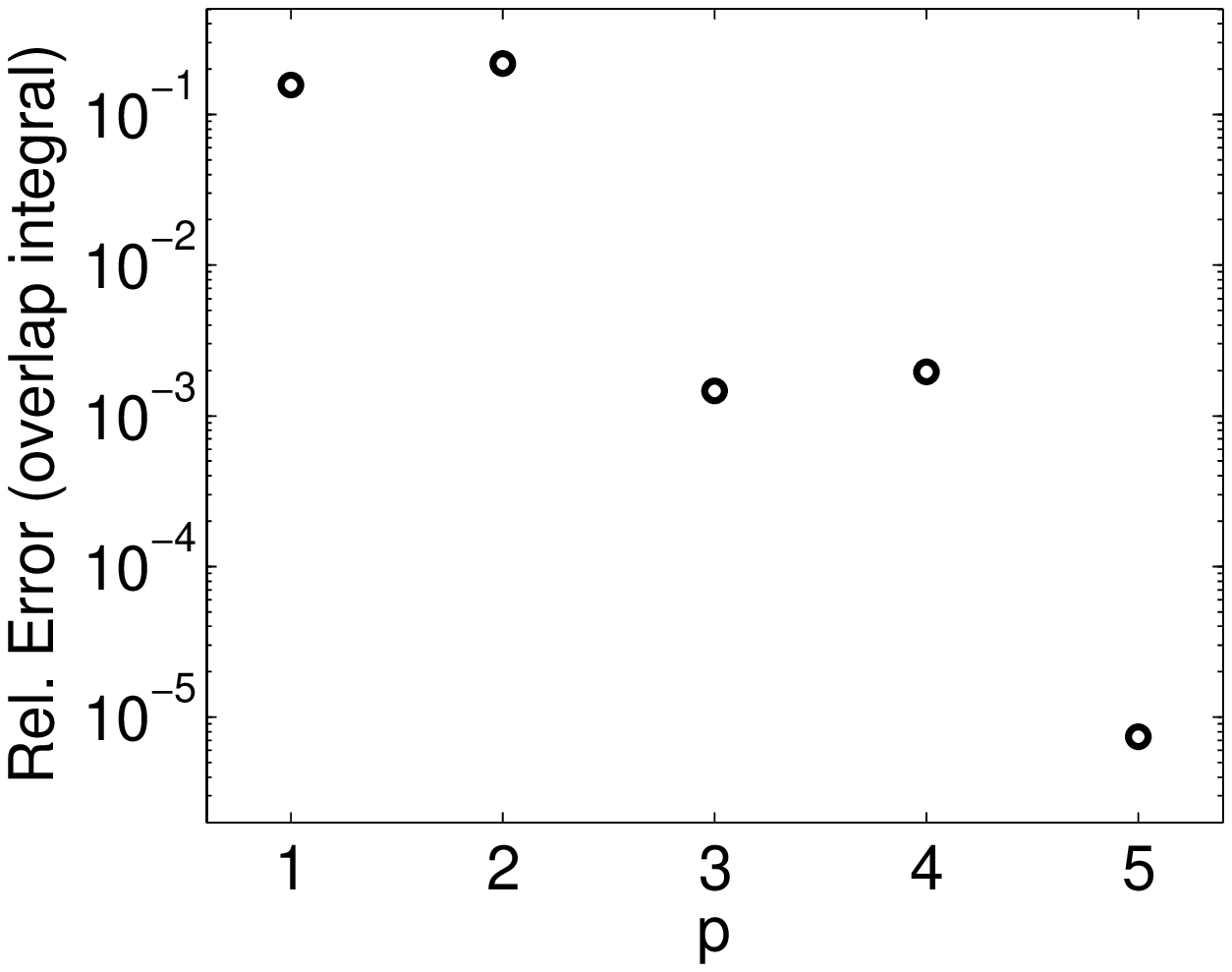}}
\caption{a) Convergence of the relative error of the total electromagnetic field energy computed 
from FEM solutions with varied finite-element degree $p$, resulting in 
computation times $T$. b) Convergence of the relative error of the overlap integral of the vertically radiated field and 
a reference field. Data obtained from the same simulations as in (a).}
\label{fig_convergence}
\end{figure}

Due to the relatively large size of the 3D computational domain and possible resonance / interference effects, 
it can be challenging to obtain 
field solutions which are accurate enough for reliable optimization~\cite{Maes2013oe}.
However, using finite-elements of higher polynomial order $p$  allows to obtain 
sufficiently accurate results within resonable computation times. 
The simulation flow is as follows: 
A wavelength scan is performed for the wavelength range of interest. 
For each wavelength $\lambda_0$: 

\begin{itemize}
\item
the waveguide modes of the waveguides entering the computational domain from the side are computed,
\item
a specific mode is defined as source field for a light scattering problem for the full 3D computational domain containing 
the grating coupler and the incoming waveguides (see, e.g., Fig.~\ref{fig:fig2} left),
\item 
the full 3D near field field distribution is computed as solution to above scattering problem, 
\item
the electromagnetic near field distribution is exported in a plane parallel to the grating coupler surface, located $1.5\,\mu$m 
above the grating surface,
\item 
overlap integrals between above scattered light field and the light fields corresponding to fundamental and higher order 
fiber modes are computed.
\end{itemize}

\begin{figure}[b]
\centering
\subfloat[]{\label{fig_fields}
\includegraphics[width=0.49\textwidth]{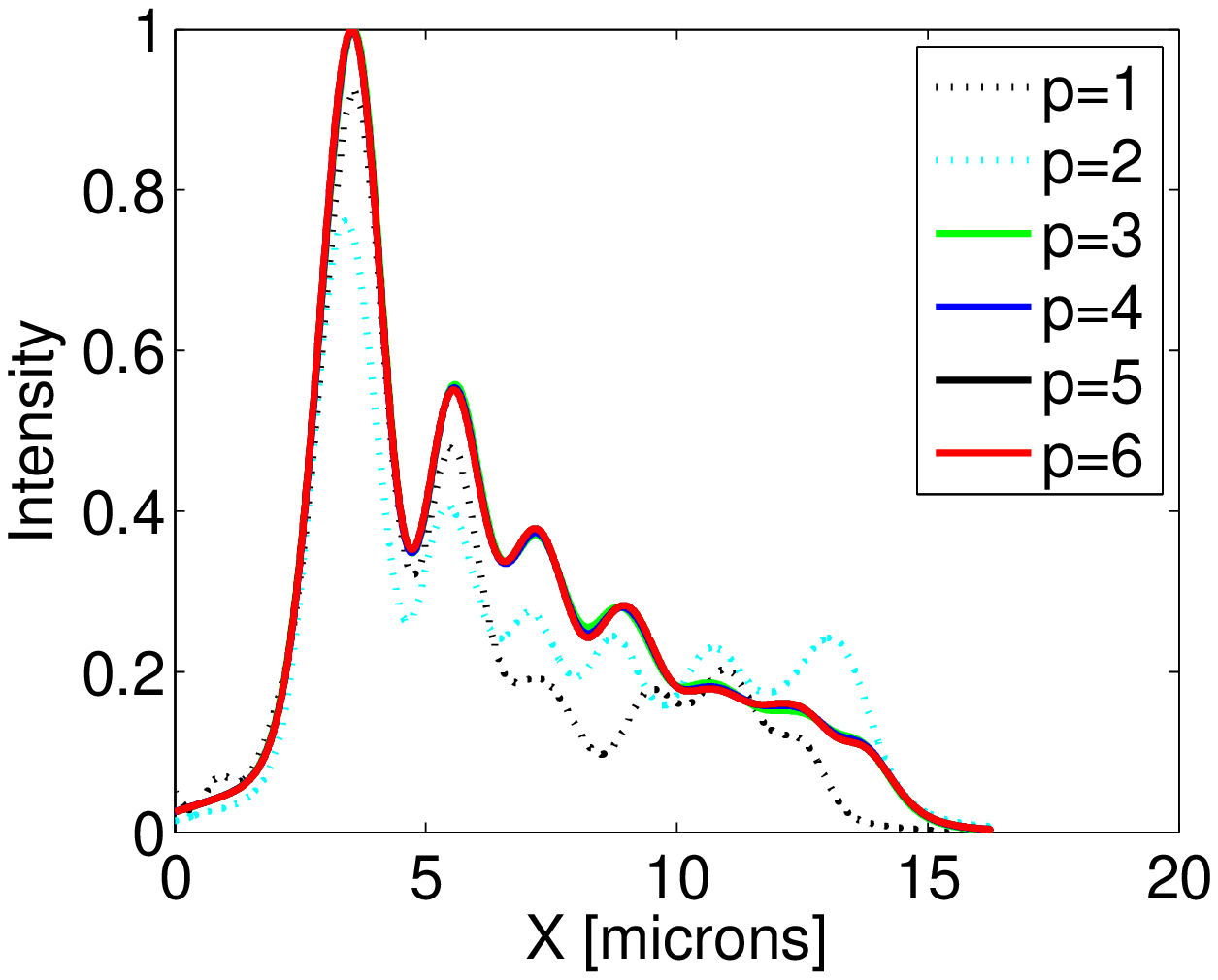}}
\hfill
\subfloat[]{\label{fig_delta_fields}
\includegraphics[width=0.49\textwidth]{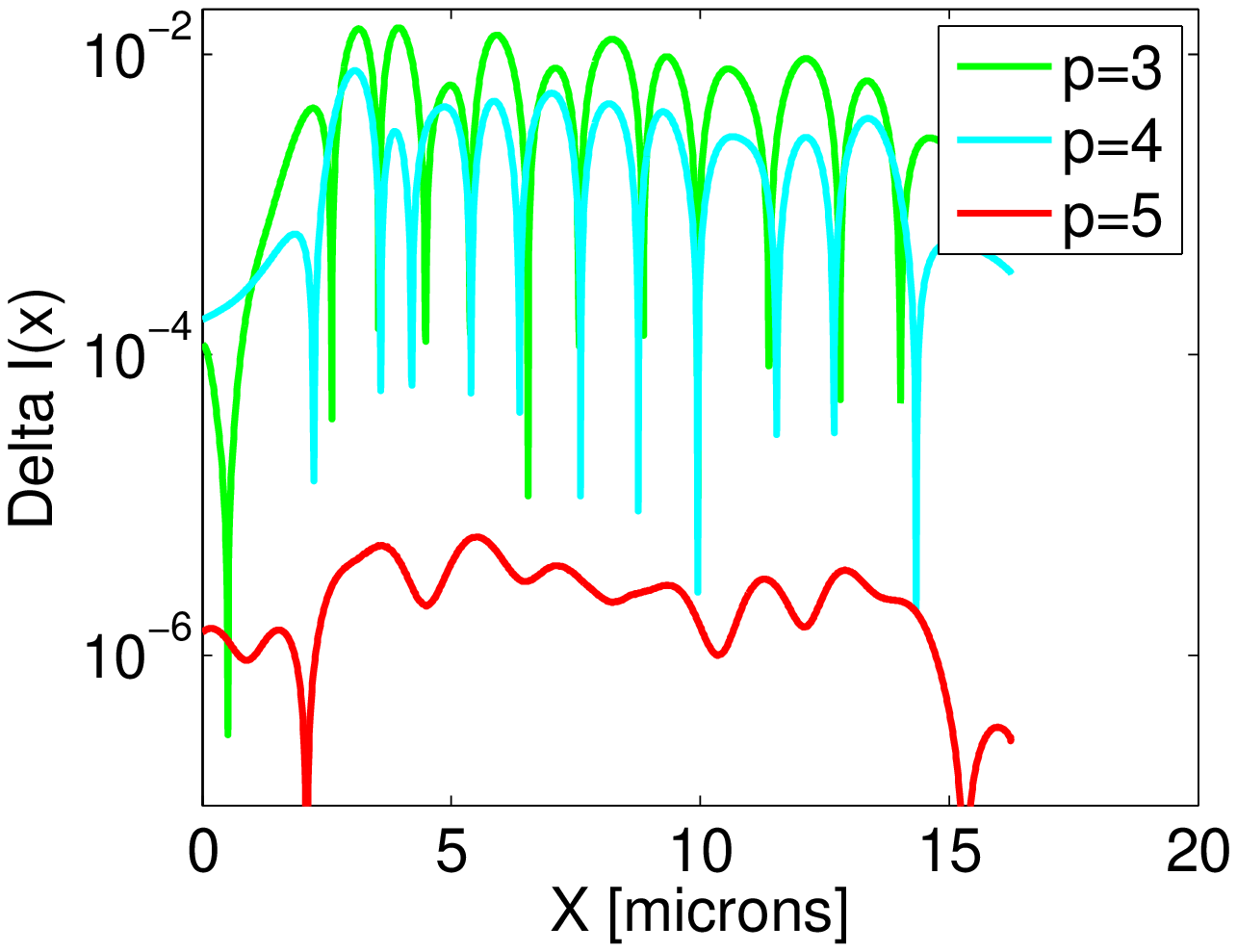}}
\caption{a) Convergence study: Field distribution above the grating coupler (where a waveguide mode is coupled in from the left), computed 
with different numerical resolution (finite-element degree $p$). b) Difference between the exported field distributions and a quasi-exact solution. }
\label{fig_field}
\end{figure}

For computation of the waveguide modes we solve an eigenvalue problem on a 2D computational domain containing the waveguide structure.
By using higher order finite-elements and adaptive mesh refinement, very high accuracies can easily be reached, such that the influence 
of numerical discretization errors in this step can be neglected~\cite{Burger2010pw1}.
Typically we choose the following numerical parameters: polynomial degree of the finite-element ansatz functions $p=4$ and 
several adaptive mesh refinement steps resulting in an unstructured spatial mesh with triangle sidelengths of sizes between 
roughly $10\,$nm and few $100\,$nm.  
For the 3D computations, a prismatoidal mesh is generated by extruding a description of the $xy$-cross-section of the device in $z$ 
direction. In each layer of the extrusion, variable domain identifiers are attributed to the extruded 3D objects. This allows 
for fast and stable construction of the 3D mesh. The mesh is constructed from prism elements, and typical sidelengths of the 
prisms in $z$-direction depend on sizes of the involved layers, up to microns. Typical sidelengths of the mesh elements
 in the $xy$-plane depend on (and are smaller than) 
the dimensions of the etched structures of few $100\,$nm.
Automatic, error-controlled mesh refinement in $z$-direction, performed by the FEM solver and depending on chosen parameter $p$ yields 
correspondingly fine 
meshes also in $z$-direction. 
For the study presented in this paper we have choosen the numerical parameter polynomial degree of the finite-element ansatz 
functions $p$ as $p=3$. We have also tested (not shown here) to choose different finite-element degrees $p$ in horizontal and 
vertial direction, and found that performance can be improved by using higher degrees in $z$-direction. 

The choice of the numerical parameters is validated in a convergence study: 
For this aim we have varied finite-element degree $p$ and observed quantitative impact on the simulation results. 
The model problem for the convergence study consists of a 3D grating with square-shaped etched pits  
($350\,\textrm{nm} \times 350\,\textrm{nm} \times 150\,\textrm{nm}$) at a periodicity length
of $p_x = p_y = 570\,$nm.
In propagation direction ($x$) of the incoming waveguide, 
the grating has 21~pits, while in the transveral direction ($y$) infinite periodicity is assumed, such 
that the size of the computational domain in $y$-direction is $p_y$ with appropriate boundary conditions. 
The total computational domain, including incoming waveguides, substrate, buffer layer (2\,$\mu$m), grating structure and air superspace 
is $16.5\,\mu\textrm{m}\times 0.57\,\mu\textrm{m}\times 4.1\,\mu\textrm{m}$, corresponding to roughly 200 cubic wavelengths at 
vacuum wavelength of $\lambda_0=1.55\,\mu$m.

Figure~\ref{fig_convergence} shows convergence of the results with varied finite-element degree $p$:
in Fig.~\ref{fig_convergence}\,a the relative error of the total field energy, $\Delta E_f$, is  plotted 
for various settings of $p$ in dependence of total computation time (CPU time on a standard workstation). 
The total field energy is the integral over the electromagnetic field energy density, integrated 
over the volume of the computational domain. The relative error is in relation to the value computed from the FEM solution 
at highest numerical resolution (at $p=6$, also called {\it quasi-exact} solution). 
Figure~\ref{fig_convergence}\,b shows convergence of the overlap integral of the intensity distribution exported to a cartesian grid 
with $\sim 60\times 1600$ points with the intensity distribution exported from the quasi-exact solution. 
This is the quantity of interest for the optimization of the grating geometrical parameters. From the Figure it can be 
seen that relative numerical errors well below 1\% are reached for $p>2$. This error level is sufficient for performing optimizations 
where the quantity of interest varies by significantly larger amounts. 

Figure~\ref{fig_field}\,a shows the field distributions for the solutions at different numerical resolution in $x$-direction, 
above the grating, integrated in $y$-direction. As expected from the convergence results above, the field distributions are quite stable 
for  $p>2$. Figure~\ref{fig_field}\,b shows the deviations of the same distributions from the quasi-exact solution. 
For  $p>2$ pointwise deviations are well below 1\%. 
In summary, the convergence study shows that the numerical error for this problem type is well controllable and that 
numerical parameter regimes can easily be found which allow for fast and sufficiently accurate performance of the solver in 
optimization loops. 
In principle, performance can further be improved by, e.g., using reduced basis methods~\cite{Pomplun2010siam} or / and automatic computation 
of parameter derivatives~\cite{burger2013al}.

%% file: pw14_wohlfeil_arxiv.bbl
\begin{thebibliography}{10}

\bibitem{salsi11}
Salsi, M., Koebele, C., Sperti, D., Tran, P., Brindel, P., Mardoyan, H., Bigo,
  S., Boutin, A., Verluise, F., Sillard, P., Bigot-Astruc, M., Provost, L.,
  Cerou, F., and Charlet, G., ``Transmission at 2x100{G}b/s, over two modes of
  40km-long prototype few-mode fiber, using {LCOS} based mode multiplexer and
  demultiplexer,'' in [{\em Optical Fiber Communication Conference/National
  Fiber Optic Engineers Conference 2011}{\nolinebreak\hspace{0.1em}]},   PDPB9,
  Optical Society of America (2011).

\bibitem{klaus12}
Klaus, W., Sakaguchi, J., Puttnam, B., Awaji, Y., Wada, N., Kobayashi, T., and
  Watanabe, M., ``Free-space coupling optics for multicore fibers,'' {\em
  Photonics Technology Letters, IEEE}~{\bf 24},  1902--1905 (2012).

\bibitem{doerr11}
Doerr, C.~R., Fontaine, N., Hirano, M., Sasaki, T., Buhl, L., and Winzer, P.,
  ``Silicon photonic integrated circuit for coupling to a ring-core multimode
  fiber for space-division multiplexing,'' {\em 37th European Conference and
  Exposition on Optical Communications} ,  Th.13.A.3, Optical Society of
  America (2011).

\bibitem{fontaine12}
Fontaine, N.~K., Doerr, C.~R., Mestre, M.~A., Ryf, R., Winzer, P., Buhl, L.,
  Sun, Y., Jiang, X., and Lingle, R., ``Space-division multiplexing and
  all-optical {MIMO} demultiplexing using a photonic integrated circuit,'' {\em
  Optical Fiber Communication Conference} ,  PDP5B.1, Optical Society of
  America (2012).

\bibitem{Burger2008ipnra}
Burger, S., Zschiedrich, L., Pomplun, J., and Schmidt, F., ``{JCMsuite}: {A}n
  adaptive {FEM} solver for precise simulations in nano-optics,'' in [{\em
  Integrated Photonics and Nanophotonics Research and
  Applications}{\nolinebreak\hspace{0.1em}]},   ITuE4, Optical Society of
  America (2008).

\bibitem{Pomplun2007pssb}
Pomplun, J., Burger, S., Zschiedrich, L., and Schmidt, F., ``Adaptive finite
  element method for simulation of optical nano structures,'' {\em phys. stat.
  sol. (b)}~{\bf 244},  3419 (2007).

\bibitem{Maes2013oe}
Maes, B., Petr\'{a}\v{c}ek, J., Burger, S., Kwiecien, P., Luksch, J., and
  Richter, I., ``Simulations of high-{Q} optical nanocavities with a gradual
  {1D} bandgap,'' {\em Opt. Express}~{\bf 21},  6794 (2013).

\bibitem{Burger2010pw1}
Burger, S., Zschiedrich, L., Pomplun, J., and Schmidt, F., ``Finite element
  method for accurate {3D} simulation of plasmonic waveguides,'' {\em Proc.
  SPIE}~{\bf 7604},  76040F (2010).

\bibitem{Pomplun2010siam}
Pomplun, J. and Schmidt, F., ``Accelerated a posteriori error estimation for
  the reduced basis method with application to 3{D} electromagnetic scattering
  problems,'' {\em SIAM J. Sci. Comput.}~{\bf 32},  498--520 (2010).

\bibitem{burger2013al}
Burger, S., Zschiedrich, L., Pomplun, J., Schmidt, F., and Bodermann, B.,
  ``Fast simulation method for parameter reconstruction in optical metrology,''
  {\em Proc. SPIE}~{\bf 8681},  868119 (2013).

\end{thebibliography}
